\documentclass[english,prb,aps,preprint]{revtex4-1}

\usepackage{amsmath}
\usepackage{amssymb}
\usepackage{graphicx}
\usepackage{babel}
\usepackage{array}
\usepackage{verbatim}
\usepackage[colorlinks=true, pdfstartview=FitV, linkcolor=blue, citecolor=blue, urlcolor=blue]{hyperref} 

\def\N        {{$^{14}$N \/}}

\def\etal     {{\it et al.}}

\newcolumntype{L}[1]{>{\raggedright\arraybackslash}p{#1}}
\newcolumntype{C}[1]{>{\centering\arraybackslash}p{#1}}

\newcommand{\captionstyle}{\normalfont} 

\newcommand{\ee}[1]{\cdot10^{#1}}
\newcommand{\mr}[1]{\mathrm{#1}}
\newcommand{\unit}[1]{\,\mathrm{#1}}

\newcommand{\oC}{^\circ{\rm C}}
\newcommand{\degree}{^\circ}

\newcommand{\cNV}{[\mr{NV^-}]}
\newcommand{\NV}{\mr{NV^-}}

\newcommand{\NVm}{\mr{NV_{m_S}^-}}
\newcommand{\NVo}{\mr{NV^0}}

\newcommand{\cN}{[\mr{N^0}]}
\newcommand{\No}{\mr{N^0}}

\newcommand{\Nm}{\mr{N_{m_S}^0}}
\newcommand{\Np}{\mr{N^+}}

\newcommand{\NVtot}{[\mr{NV_{tot}}]}

\newcommand{\yN}{\Gamma_\mr{N^0}}
\newcommand{\yNV}{\Gamma_\mr{NV^-}}
\newcommand{\ko}{k_0}
\newcommand{\kA}{k_A}
\newcommand{\kB}{k_B}
\newcommand{\kC}{k_C}
\newcommand{\kD}{k_D}

\newcommand{\pNVa}{p_\mr{NV_{-1}^-}}
\newcommand{\pNVb}{p_\mr{NV_{0}^-}}
\newcommand{\pNVc}{p_\mr{NV_{+1}^-}}
\newcommand{\pNVj}{p_\mr{NV_j^-}}

\newcommand{\pNV}{p_\mr{NV^-}}
\newcommand{\pNVo}{p_\mr{NV^0}}

\newcommand{\pNa}{p_\mr{N_{-1/2}^0}}
\newcommand{\pNb}{p_\mr{N_{+1/2}^0}}
\newcommand{\pNj}{p_\mr{N_j^0}}

\newcommand{\pN}{p_\mr{N^0}}
\newcommand{\pNo}{p_\mr{N^0}}
\newcommand{\pNp}{p_\mr{N^+}}

\newcommand{\PNV}{P_\mr{NV^-}}
\newcommand{\PN}{P_\mr{N^0}}

\begin{document}

\title{Optical hyperpolarization of nitrogen donor spins in bulk diamond}

\author{M. Loretz$^1$, H. Takahashi$^1$, T. F. Segawa$^{1,2}$, J. M. Boss$^1$, C. L. Degen$^1$}
\affiliation{$^1$Department of Physics, ETH Zurich, Otto Stern Weg 1, 8093 Zurich, Switzerland.}
\affiliation{$^2$Department of Chemistry and Applied Biosciences, ETH Zurich, Vladimir Prelog Weg 2, 8093 Zurich, Switzerland.}

\email{degenc@ethz.ch}

\begin{abstract}
We report hyperpolarization of the electronic spins associated with substitutional nitrogen defects in bulk diamond crystal.  Hyperpolarization is achieved by optical pumping of nitrogen vacancy centers followed by rapid cross relaxation at the energy level matching condition in a 51 mT bias field.  The maximum observed donor spin polarization is 0.9 \% corresponding to an enhancement by 25 compared to the thermal Boltzmann polarization.  A further accumulation of polarization is impeded by an anomalous optical saturation effect that we attribute to charge state conversion processes.  Hyperpolarized nitrogen donors may form a useful resource for increasing the efficiency of diamond-based dynamic nuclear polarization devices.
\end{abstract}

\date{\today}

\maketitle

\section{Introduction}

Nuclear magnetic resonance (NMR) spectroscopy is a powerful method for contemporary molecular characterization and diagnostics due to a superb spectroscopic resolution.  By contrast, the sensitivity of NMR is low due to its reliance on the Boltzmann polarization, which is $P\leq 10^{-4}$ even for the highest fields accessible with state-of-the-art superconducting magnets \cite{hashi15,bruker16}.  The low sensitivity posits that a sufficiently large sample volume in high enough concentration is available, which presents a significant obstacle when delicate samples and nuclei with a low isotope abundance or low gyromagnetic ratio are involved.  In order to circumvent the sensitivity limitations of NMR, hyperpolarization techniques that increase the polarization beyond the Boltzmann level -- ideally approaching 100\% polarization -- are being widely explored.  In particular, these include triplet dynamic nuclear polarization (DNP) \cite{henstra90}, dissolution DNP \cite{ardenkjaer03,gajan14}, optically pumped noble gases \cite{grover78,bhaskar82,goodson02,lilburn13} and semiconductors \cite{lampel68,tycko96,tycko98}, and many varieties of those techniques \cite{griffin10,reimer10}.  The main drawbacks of most hyperpolarization methods are a cost-expensive cryogenic hardware and an often-limited area of application.

A new concept for an inexpensive, room temperature polarizer is based on diamond crystals doped with nitrogen-vacancy defects (NV centers) \cite{drake16,fischer13,scheuer16,abrams14,king15}.  In a diamond polarizer, hyperpolarization is induced by optical pumping of NV defect spins using intense laser illumination.  Diamond polarizers are proposed to have many advantages, which include beside the room-temperature compatibility, rapid optical pumping, near-unity polarization, and no need for sample doping by radicals.  In addition, the chemical inertness and stability of diamond makes the device potentially reusable for a large variety of liquid samples. 
The central bottleneck of diamond polarizers, however, is the transport of polarization from the defect spins within diamond to target spins outside of diamond.  This step has to date only been demonstrated indirectly and on a few-spin scale \cite{mamin13,staudacher13,loretz14apl}.  Because the separation between source and target nuclear spins must be less than $\sim5\unit{nm}$, only defect centers located in the top surface layer contribute to the transfer.  Thus, even for high NV densities of order $10^{18}\unit{cm^{-3}}$ ($\sim 10\unit{ppm}$) \cite{su13} the number of actively participating spins is low.  One proposed remedy is to structure the surface so as to increase its effective surface area \cite{abrams14}.

\begin{figure}
\includegraphics[width=0.92\columnwidth]{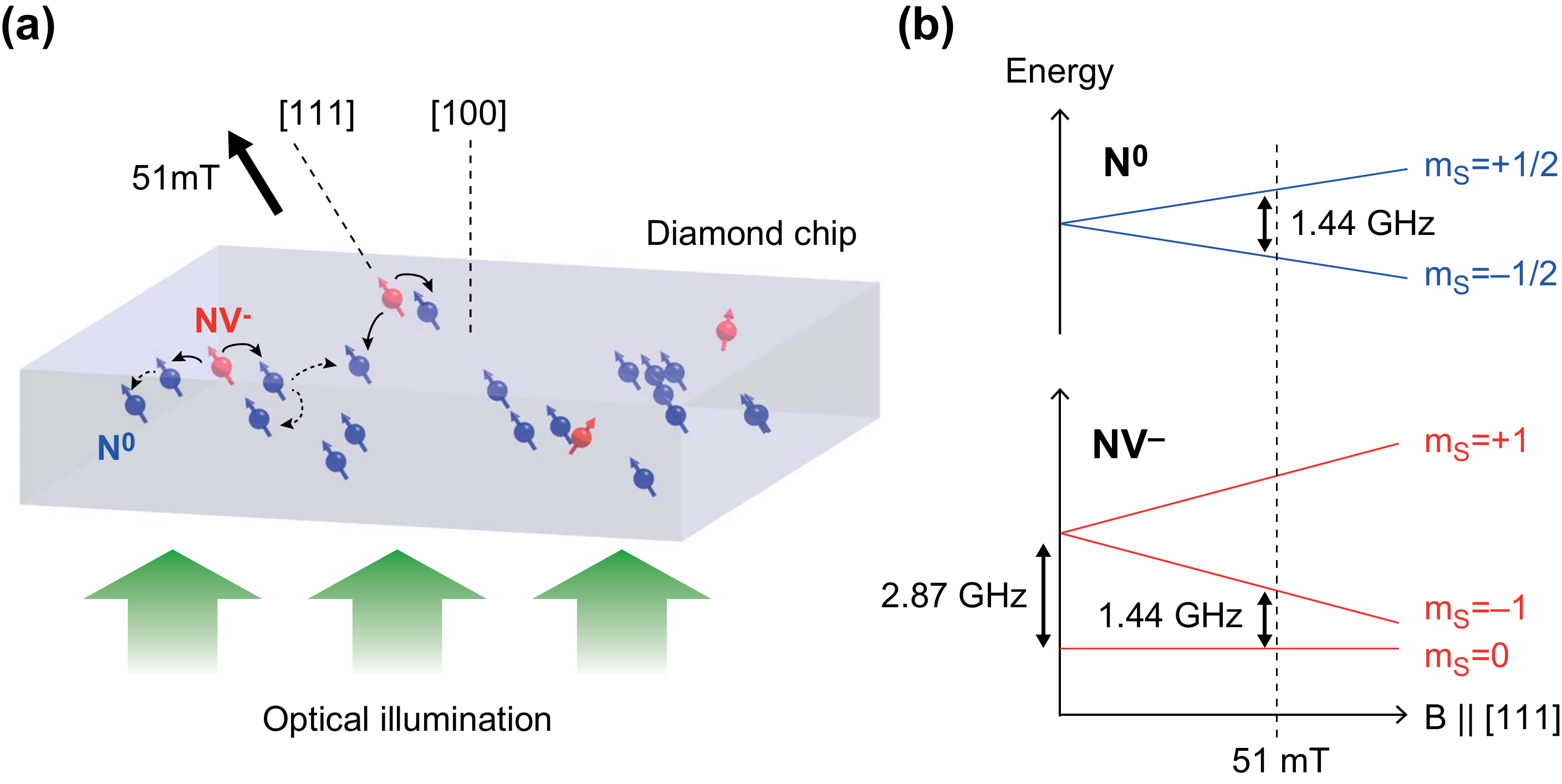}
\caption{\captionstyle
(a) Basic arrangement for optical hyperpolarization of nitrogen donor spins in diamond.
In a first step, nitrogen vacancy centers ($\NV$, red, spin $S=1$) are optically polarized into the $m_S=0$ spin state by exposure to a green laser beam \cite{doherty13}.  Next, the polarization is transferred to adjacent nitrogen defects ($\No$, blue, spin $S=1/2$) using cross-relaxation mediated by the dipolar interaction (solid arrows).  Only those $\NV$ centers whose symmetry axis is aligned with the external bias field ($B||[111]$) are optically pumped and contribute to cross-relaxation.  Finally, the polarization is rapidly distributed in the nitrogen spin bath via spin diffusion (dashed arrows).  The polarization may later be transferred to target nuclear spins outside the diamond chip.  This could be achieved either directly, using the integrated solid effect \cite{henstra14} or the Overhauser effect \cite{gunther13}, or indirectly, via additional spin labels on the diamond surface \cite{waddington16} or in the analyte.
(b) Spin energy level diagram indicating the matching condition between the $m_S=0\leftrightarrow m_S=-1$ ($\NV$) and $m_S=-\frac12\leftrightarrow m_S=+\frac12$ ($\No$) transitions at $51\unit{mT}$.
\label{fig:arrangement}
}
\end{figure}

Another potential route for increasing hyperpolarization efficiency, explored here, takes advantage of abundant substitutional nitrogen defects (P1 centers) \cite{kaiser59,smith59}.  Rather than directly polarizing an outside analyte, the polarization is first transferred to the large bath of nitrogen donor spins using a cross-relaxation process (see Fig. \ref{fig:arrangement}).  Since P1 centers can approach densities of $10^{20}\unit{cm^{-3}}$ ($\sim 400\unit{ppm}$) \cite{su13}, the electronic spin polarization at the diamond surface can potentially be greatly increased.  In addition, because P1 centers, in contrast to NV centers, do not require a preferential alignment with the external bias field, they are more suited for transferring polarization to outside nuclear spins.

In this paper, we investigate hyperpolarization of nitrogen donor spins in bulk diamond by \textit{in situ} electron paramagnetic resonance (EPR) spectroscopy.  The hyperpolarization is induced by optical pumping of NV centers and cross-relaxation at the energy level matching of the two spin species at $B=51\unit{mT}$ (see Fig. \ref{fig:arrangement}b).  We observe that large polarization enhancement factors, exceeding 100 for NV centers and up to 25 for P1 centers, can be generated.  In addition, we observe that the polarization enhancement unexpectedly saturates already at very low laser intensities, and far below the known saturation intensity for NV centers.  By comparing the results to a kinetic model of spin populations we find that the anomalous saturation can be partially explained by charge state conversions of NV and P1 defects.  To our knowledge, this study is the first demonstration of nitrogen donor hyperpolarization in the bulk.
%


\section{Experimental}

Experiments are carried out on a home-built EPR spectrometer specifically designed for operation at $51\unit{mT}$.  The EPR probe consists of an inductively coupled loop gap resonator resonant at $1.44\unit{MHz}$ with a quality factor of 1,500 [\onlinecite{rinard05},\onlinecite{chen88}].  The magnetic field is provided by a pair of permanent disc magnets arranged in a Helmholtz configuration and an additional Helmholtz coil with a sweep range of $4\unit{mT}$.  In order to orient the magnetic field along one of the {[}111{]} diamond crystal axes the resonator is placed at an angle of $35\degree$ with respect to the magnetic field.  The diamond crystal is then rotated by approximately $45\degree$ around the longitudinal axis of the resonator until the desired orientation is reached.  EPR spectra are acquired with a standard continuous-wave detection scheme with a field modulation of $9.7\unit{kHz}$ and a low microwave power of a few $100\unit{nW}$.  To optically pump the NV centers, a laser beam is directed through an opening along the longitudinal direction of the resonator and onto the flat (100) surface of the diamond chip.  The laser beam has a diameter of $\sim 5\unit{mm}$ corresponding to an intensity of $\sim 50\unit{mW/mm^2}$ for 1 W of incident light.  The optical beam is generated by a $532\unit{nm}$ cw laser (Coherent Verdi 10 W) and gated by a shutter.  To mitigate the absorptive heating the diamond chip is exposed to a flow of cold nitrogen gas ($\sim200$ K) via a central sample tube.

Two diamond chips are used in this study.  Both chips are grown by high-pressure, high-temperature synthesis and are of type Ib (ElementSix Ltd.).  The chips have exposed (100) surfaces and lateral dimensions of $3\times3\unit{mm^2}$ and a thickness of $0.3\unit{mm}$ (chip A) and $0.08\unit{mm}$ (chip B), respectively.  To increase the NV density, chip A (chip B) is irradiated by a $2\unit{MeV}$ electron beam (Leoni Studer AG) for a total duration of $30\unit{h}$ ($40\unit{h}$) with an intermediate annealing step after $10\unit{h}$ ($20\unit{h}$) and a final annealing step at the end.  Annealing is performed in high vacuum and at $850\oC$ for $2\unit{h}$ \cite{acosta09,su13}.  A wide-field fluorescence measurement is used to confirm that the NV density is uniform over the chip surface.  Defect densities of the relevant $\NV$ and $\No$ charge states are quantified by cw EPR spectroscopy on a Bruker ElexSys E500 X-Band (9.6 GHz) spectrometer by comparing the double-integrated signal intensities with a standard sample.  Defect densities are $\cN = 77\unit{ppm}$ and $\cNV = 9\unit{ppm}$ for chip A as well as $\cN = 36\unit{ppm}$ and $\cNV = 11\unit{ppm}$ for chip B (see Table \ref{tab:enhancement}) with an uncertainty of $20\%$.


\section{Results}

\subsection{EPR spectroscopy at 51 mT}

\begin{figure}
\includegraphics[width=0.80\columnwidth]{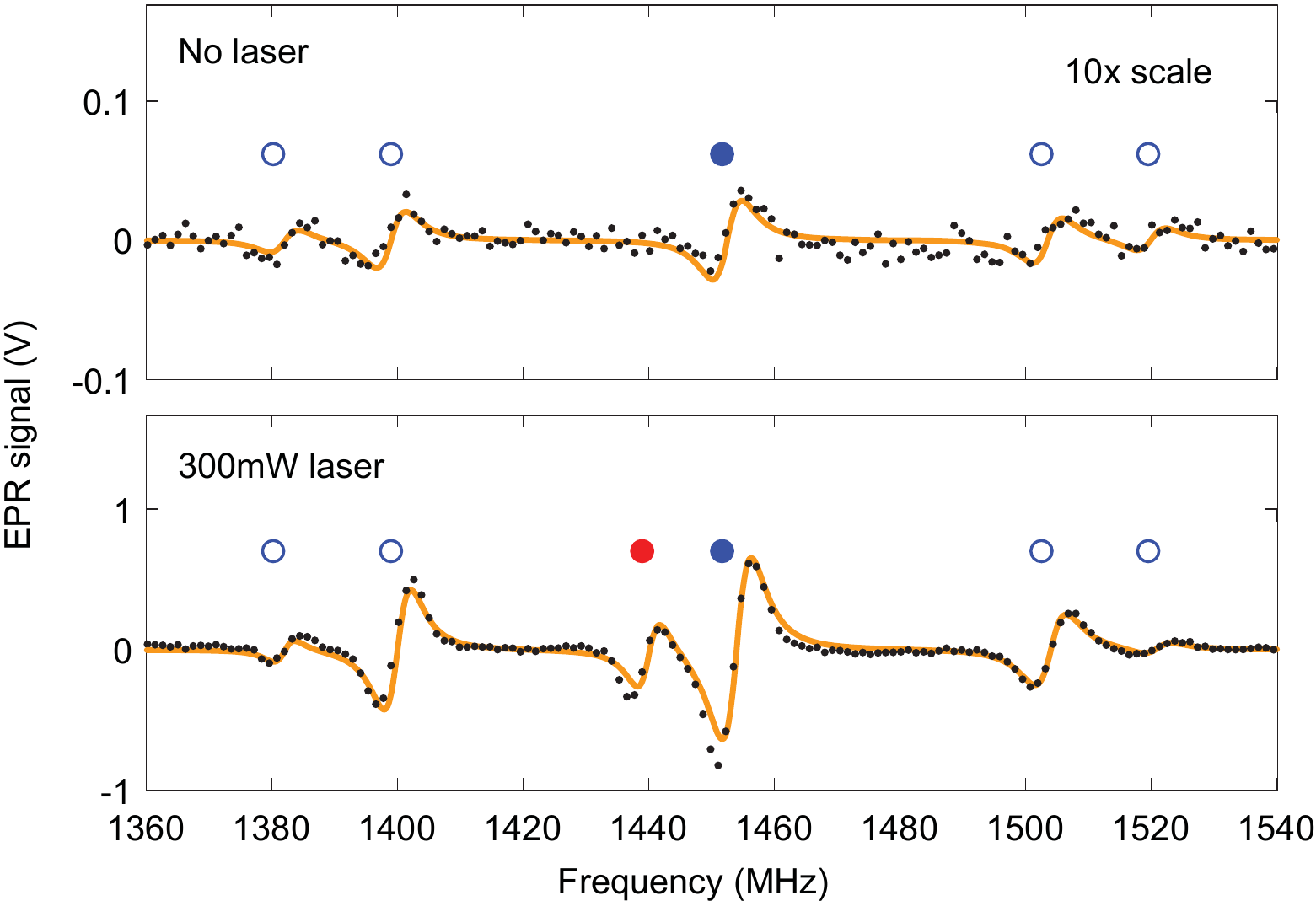}
\caption{\captionstyle
(a) EPR spectrum at $51\unit{mT}$ of diamond chip A with no laser illumination.  Five $\No$ resonances are visible, including the central transition (filled blue circle) and the four hyperfine-shifted transitions (open blue circles) associated with the $I=1$ nitrogen \N nuclear spin.  The $\NV$ resonance is below the detection limit.
(b) EPR spectrum with $300\unit{mW}$ laser illumination.  The $m_S=0\leftrightarrow m_S=-1$ transition of aligned $\NV$ centers appears as an additional peak at $\sim 1440\unit{MHz}$ (filled red circle); other $\NV$ resonances fall outside of the spectral range.  A strong polarization enhancement is observed for both $\NV$ and $\No$.  Solid lines are fits.  Spectra are recorded in a single passage with a dwell time of about 1 s per point.  Units have been converted from field to frequency using the free-electron $g$-factor.  Additional quantities are given in Table \ref{tab:enhancement}.}
\label{fig:eprspectra}
\end{figure}

Figure \ref{fig:eprspectra} shows two representative EPR spectra recorded with the home-built setup at $51\unit{mT}$ in the absence and presence of green laser illumination.  Without illumination, the spectrum shows the typical pattern of five resonances centered around $1452\unit{MHz}$ caused by the center and \N hyperfine satellite transitions of $\No$ (Fig. \ref{fig:eprspectra}(a)) [\onlinecite{smith59}].  The $\NV$ resonance is not visible in this spectrum because the Boltzmann polarization is too low to produce a detectable signal at these low concentrations.

Under optical illumination with $300\unit{mW}$ the $m_S=0\leftrightarrow m_S=-1$ resonance of the $\NV$ signal becomes visible at $\sim 1440\unit{MHz}$ (see Fig. \ref{fig:eprspectra}(b)).  In addition, the five $\No$ resonances are strongly enhanced.  This demonstrates that both spin species are hyperpolarized and that the cross relaxation mechanism is effective.  Despite of the fact that the level matching condition is only fulfilled for the central $\No$ resonance, a similar polarization enhancement is observed for all four hyperfine-shifted $\No$ resonances, probably because of spectral diffusion due to nuclear spin flips.  To evaluate the polarization enhancement, we can fit each resonance by the derivative of a Gaussian function and subsequently double integrate and add all resonances.  For the non-illuminated spectrum, the integrated signal intensity is $11\unit{a.u.}$ for $\No$.  Considering the proportions of defect densities (Table \ref{tab:enhancement}), the corresponding $\NV$ intensity is estimated at $0.32\unit{a.u.}$  Under optical illumination, the signal intensities are $280\unit{a.u.}$ for $\No$ and $33\unit{a.u.}$ for $\NV$.  The corresponding polarization enhancement is therefore about $100$ and $25$, respectively (see Table \ref{tab:enhancement}).  Because the signal in the non-illuminated reference spectrum is small, the uncertainty in these enhancement factors is rather large, on the order of 25\% (2 s.d.).
\begin{table}
\begin{tabular}{lC{1.5cm}C{1.5cm}C{2cm}C{2cm}C{2.5cm}C{2.5cm}c}
\hline\hline
Defect & Charge state & Spin    & Density & Effective density & Polarization (Boltzmann) & Polarization (300 mW) & Enhancement \\
\hline 
NV     & $\NV$        & $S=1$   & 9 ppm   & 2.3 ppm           & 0.034 \%             & 3.5\%                  & 103 \\  
P1     & $\No$        & $S=1/2$ & 77 ppm  & 77 ppm            & 0.034 \%             & 0.86\%                  & 25 \\
\hline\hline
\end{tabular}
\caption{\captionstyle Important quantities for the measurements in Fig. \ref{fig:eprspectra}.  Effective $\NV$ density and polarizations refer to the $\sim 25\%$ of NV centers whose symmetry axes are aligned with the bias field.  Boltzmann polarization is reported for 200 K.}
\label{tab:enhancement}
\end{table}

\begin{figure}
\includegraphics[width=0.80\columnwidth]{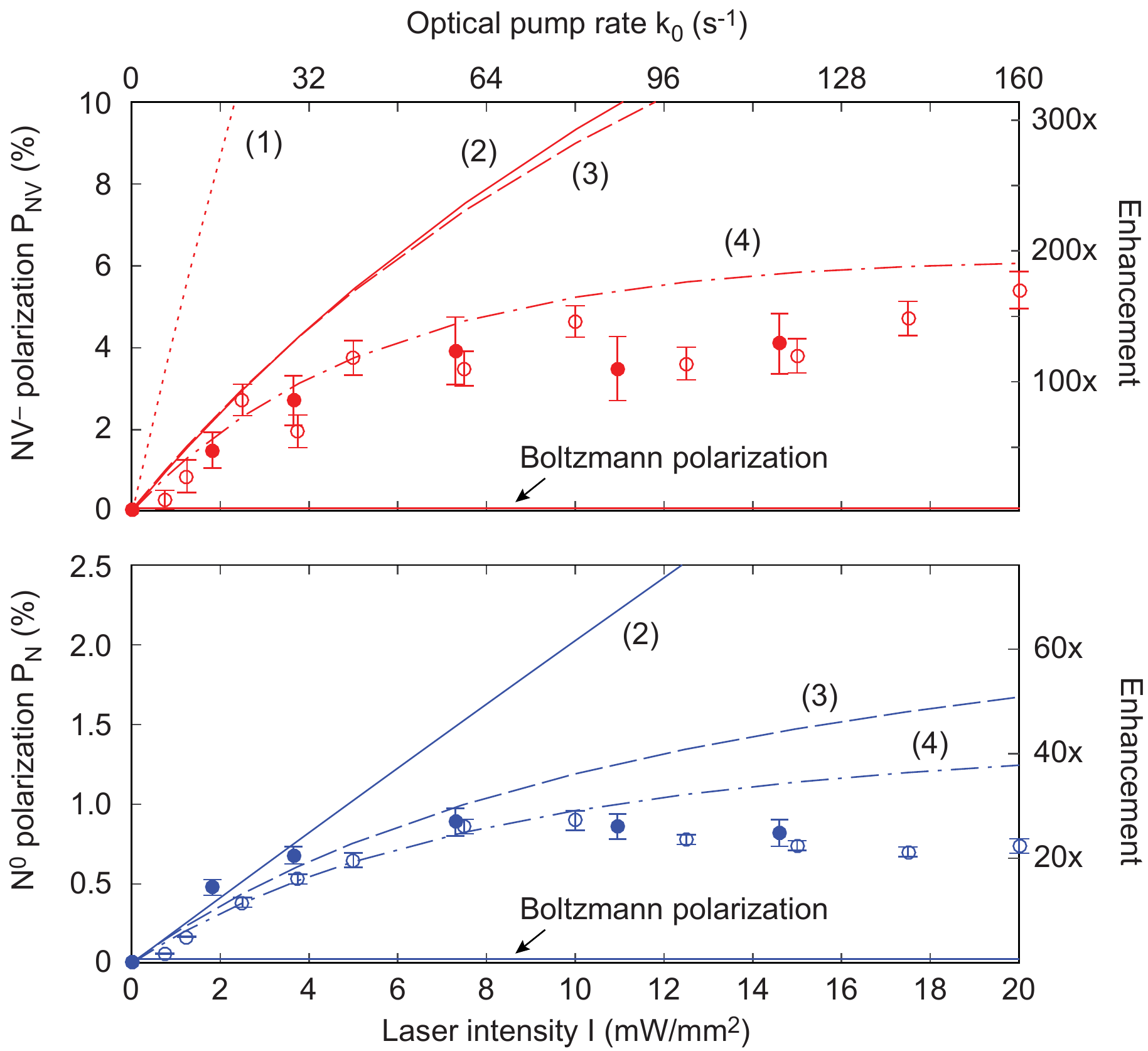}
\caption{\captionstyle Spin polarization of $\NV$ and $\No$ plotted as function of laser intensity.  The corresponding laser powers are $0-800\unit{mW}$.  Open circles represent chip A and closed circles represent chip B. 
The maximum observed polarization enhancement is 170 for $\NV$ and 25 for $\No$.
Solid and dashed lines represent the different kinetic models discussed in the main text: (1) No cross relaxation, $\PNV=\ko/[\ko+3\yNV]$, (2) Simple cross relaxation, (3) Cross relaxation and $\No\leftrightarrow\Np$ charge state conversion, (4) Cross relaxation and $\No\leftrightarrow\Np$, $\NV\leftrightarrow\NVo$ charge state conversions.  An anomalous saturation effect is observed for laser intensities above ca. $10\unit{mW/mm^2}$ which is not explained by the models.  Model parameters are given in Table \ref{tab:model} and are for chip A; the slight deviation of the chip B data for the $\No$ panel is due to the different $\NV$ concentration.  Error bars are fit errors (2 s.d.).
}
\label{fig:eprlaserpower}
\end{figure}

To investigate the maximum possible polarization enhancement we measure the induced spin polarizations as a function of laser power.  Fig. \ref{fig:eprlaserpower} plots the spin polarization for $\NV$ and $\No$ versus laser intensity.  The spin polarization is calculated by comparing the EPR signal intensities to the reference values from Table \ref{tab:enhancement}.  The laser intensity $I$ represents an effective intensity because the deeper regions of the diamond chips receive less light due to optical absorption; the reduction compared to the incident intensity is approximately $0.50$ for chip A and $0.73$ for chip B, respectively, as inferred from the absorption cross section of the $\NV$ center ($\sigma = 3\ee{-17}\unit{cm^2}$, Ref. \onlinecite{wee07}) and the chip thicknesses.
Fig. \ref{fig:eprlaserpower} shows that at low laser intensities, $I\lesssim 4\unit{mW/mm^2}$, the observed spin polarization increases proportionally with the optical power.  This behavior is expected in the regime where the optical pump rate $\ko$ is slower than the spin relaxation rate $\yNV=(T_{1,\mr{\NV}})^{-1}$.  The pump rate can be estimated from the absorption cross section of the $\NV$ center and is about $\ko = 8\ee{3}\unit{s^{-1}}\times I/[\mr{W mm^{-2}}]$ (Ref. \onlinecite{wee07}).  This yields $\ko \approx 32\unit{s^{-1}}$ for $I=4\unit{mW/mm^2}$ (see upper scale in Fig. \ref{fig:eprlaserpower}) which is indeed below the expected spin relaxation rate of $\yNV\sim 50\unit{s^{-1}}$ [\onlinecite{jarmola12}].

By contrast, at laser intensities above $10\unit{mW/mm^2}$ the polarization saturates and does not exceed $\sim 4\%$ for $\NV$ and $\sim 0.6\%$ for $\No$, respectively.  This saturation is unexpected because the optical intensity is orders of magnitude below the optical saturation limit for $\NV$ centers ($\sim 10^3\unit{W/mm^2}$)  [\onlinecite{chen15guo}].  The saturation observed in Fig. \ref{fig:eprlaserpower} is therefore not related to a simple electronic excitation of the NV center and must be caused by another mechanism.  A similar anomalous saturation has recently been reported in an EPR study by Drake \etal \cite{drake16}, which was carried out at X-band at $9.7\unit{GHz}$ where cross relaxation can be neglected.  Drake \etal\ observed a saturation of the EPR signal for laser intensities between $10-20\unit{mW/mm^2}$, in good agreement with our finding.  In the following we argue, based on a simple kinetic model, that the anomalous saturation effect observed in both Ref. [\onlinecite{drake16}] and our study is probably due to unspecified charge state conversion processes.

\subsection{Modeling of polarization dynamics including cross relaxation}

\begin{figure}[h]
\includegraphics[width=0.75\columnwidth]{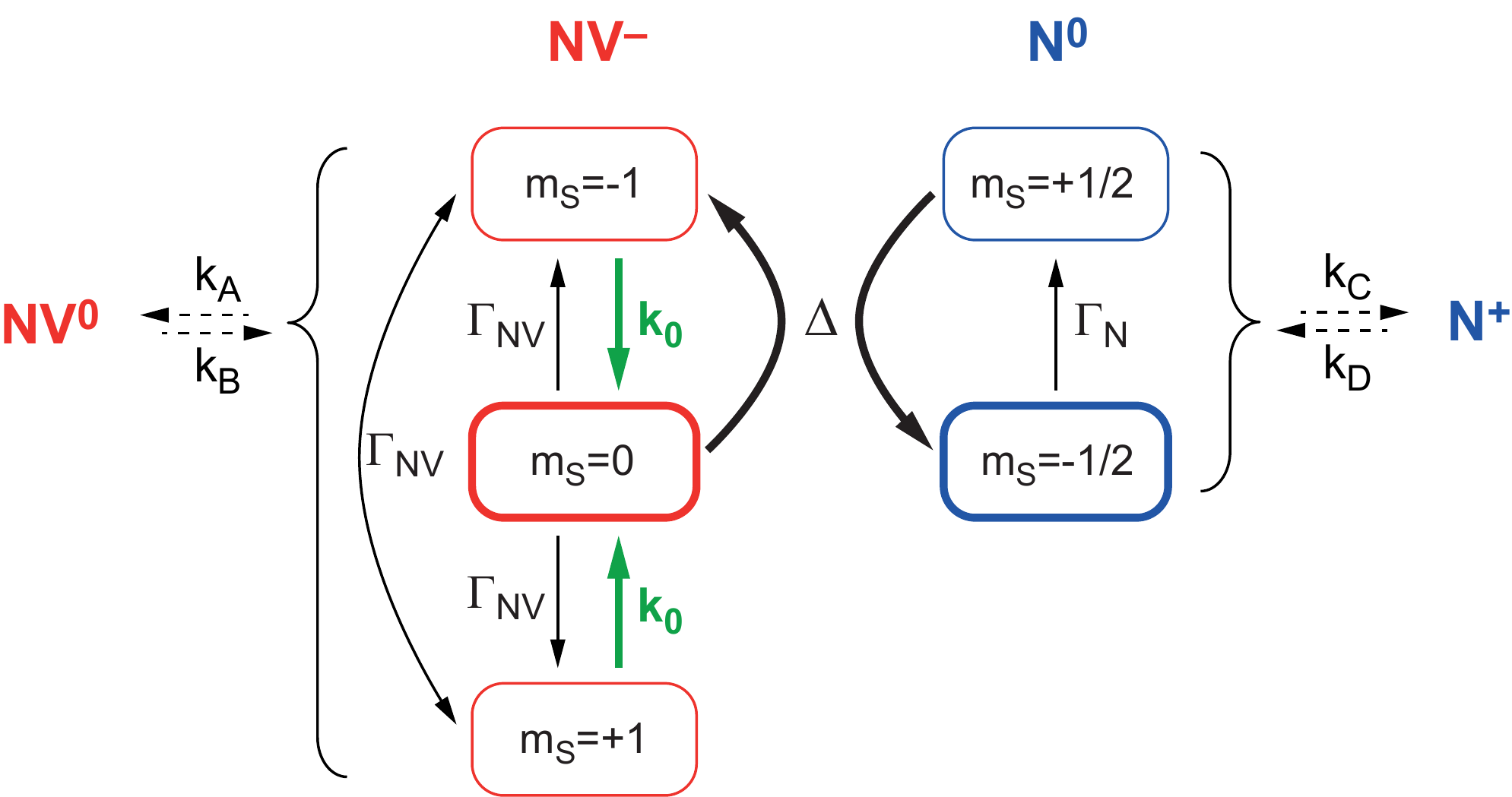}
\caption{\captionstyle Reservoir model of the NV-P1 system.
Green bold arrows indicate the pumping process, black bold arrows indicate rapid cross-relaxation, and bold boxes represent majority populations.  Dashed arrows indicate charge state conversions.  Solid arrows indicate the principal direction of population flow.  Rate constants are explained in the text.
}
\label{fig:boxschematic}
\end{figure}

We have analyzed the polarization build-up in terms of population dynamics between coupled spin baths (see Fig. \ref{fig:boxschematic}).  The most basic model takes into account longitudinal spin relaxation rates $\yNV$ and $\yN$ for $\NV$ and $\No$, respectively, and a pair-wise cross-relaxation rate $\Delta$ between the reservoirs.  Later on we will extend the model to also include charge state conversions to neutral $\NVo$ and positive $\Np$, respectively.

In a first step, we can inspect the build-up of polarization in the absence of charge state conversion processes.  The dynamics of the spin populations are then described by the following set of rate equations,
\begin{align}
\frac{d\pNVa}{dt}
  &= - \ko\pNVa
	   + \yNV(-2\pNVa+\pNVb+\pNVc)
		 - R\Delta (\pNVa\pNa-\pNVb\pNb)  \label{eq:cross:a} \\
\frac{d\pNVb}{dt}
  &= + \ko(\pNVa+\pNVc)
	   + \yNV(-2\pNVb+\pNVa+\pNVc)
		 + R\Delta (\pNVa\pNa-\pNVb\pNb) \\
\frac{d\pNVc}{dt}
  &= - \ko\pNVc
	   + \yNV(-2\pNVc+\pNVa+\pNVb) \\
\frac{d\pNa}{dt}
  &= - \yN(\pNa-\pNb)
	   - \Delta(\pNVa\pNa-\pNVb\pNb) \\
\frac{d\pNb}{dt}
  &= - \yN(-\pNa+\pNb) 
	   + \Delta(\pNVa\pNa-\pNVb\pNb) \ ,
\end{align}
where $\pNVa$, $\pNVb$ and $\pNVc$ denote the relative populations (probabilities) of the three $\NVm$ spin states and $\pNa$, $\pNb$ denote the relative populations of the two $\Nm$ spin states. The constant $R = \cN/(0.25\cNV)$ is the ratio between the sizes of the $\No$ and effective $\NV$ spin baths (see Table \ref{tab:model}).
Note that we assume equal spin relaxation rates between all three $\NV$ spin levels.  Although the relaxation between the $m_S=\pm 1$ states is likely much slower, the assumption of equal relaxation rates allows for a simple analysis.  We verified through numerical simulations that the assumption only has a minor effect on the overall result.

We have resolved this model to calculate the equilibrium polarizations in the limit of fast cross-relaxation ($\Delta \gg \yNV,\yN,\ko$).  The polarizations are
\begin{align}
\PNV &= \pNVb - \frac12(\pNVa+\pNVc) = \frac{\ko(2\ko+R\yN+4\yNV)}{(\ko+2R\yN+2\yNV)(2\ko+6\yNV)} \ ,  \label{eq:pnv:a} \\
\PN  &= \pNb - \pNa = \frac{\ko}{\ko+2R\yN+2\yNV} \ .  \label{eq:pn:a}
\end{align}
Equations (\ref{eq:pnv:a}) and (\ref{eq:pn:a}) are plotted as solid lines (Curve 2) along side the experimental data in Fig. \ref{fig:eprlaserpower} for the parameters given in Table \ref{tab:model}.  We find that the model well describes the polarization enhancement of both spin species at low laser intensities, but that it cannot account for the saturation observed for intensities above roughly $10\unit{mW/mm^2}$.

\subsection{Modeling of polarization dynamics including charge state conversions}

In a second step we analyze whether charge state conversion processes between $\NV$ and neutral $\NVo$ as well as between $\No$ and ionized $\Np$, which are known to be promoted by laser illumination \cite{manson05,aslam13}, can be responsible for the anomalous saturation.  Charge state conversions work to reduce the absolute EPR signal intensity through two effects.  First, a significant accumulation of $\NVo$ and $\Np$ depletes the sizes of the spin reservoirs.  Second, as charge state conversions are typically not spin preserving, they lead to accelerated spin relaxation.
Several mechanisms are known to convert between charge states.  The neutral charge state $\NVo$ can be induced by two-photon absorption via the $\NV$ excited state and can reconvert to $\NV$ by a similar two-photon absorption process \cite{manson05,chen15guo,bourgeois15}.  Because our optical intensity is very low, however, we can exclude a significant contribution by these processes.  Instead, we assume that the conversion from $\NV\rightarrow\NVo$ occurs via electron tunneling from a photo-excited $\NV$ to a nearby $\Np$, and the reconversion via electron capture from the conduction band \cite{manson05}.  The conversion between $\No$ and $\Np$ is assumed to occur via a simple one-photon ionization and recombination process \cite{heremans09,jayakumar16}.

To account for charge state conversions we extend the model by the following four rate equations,
\begin{align}
\frac{d\pNVj}{dt}  &= - R'\kA \pNVj \pNp + \frac13\kB n \pNVo  \label{eq:csc:a} \ , \\
\frac{d\pNVo}{dt}  &= + R'\kA \pNV \pNp - \kB n\pNVo  \label{eq:csc:b} \ , \\
\frac{d\pNj}{dt}   &= - \kC \pNj + \kD n \pNp + \frac12 \kA \pNV \pNp  \label{eq:csc:c} \ , \\
\frac{d\pNp}{dt}   &= + \kC \pNo - \kD n \pNp - \kA \pNV \pNp  \label{eq:csc:d} \ , 
\end{align}
where $\kA$ through $\kD$ are rate constants and $R'=\cN/\cNV=R/4$ (because all NV orientations participate in charge transfer).  The subscript $j$ indicates the respective spin state where $\pNV = \pNVa+\pNVb+\pNVc$ and $\pN = \pNa+\pNb$.
Rates $\kA$ and $\kC$ depend on photon absorption and are proportional to $\ko$.  $\kB$ and $\kD$ represent charge recombination and are proportional to the charge carrier density $n$, where $n$ is normalized by the NV density $\NVtot$.  If we stay within our model, neutrality of charge requires that $n = R'\pNp-\pNV$.  Under illumination, $n$ is dominated by photo-excited electrons and hence also dependent on the optical power.

We numerically evaluate the extended model described by Eqs. (\ref{eq:cross:a}-\ref{eq:csc:d}) using the Euler method for two situations.  A first case only takes $\No\leftrightarrow\Np$ charge state conversions into account but neglects $\NV\leftrightarrow\NVo$ ($\kA=\kB=0$).  The second scenario includes both types of charge state conversions.  Because the rate constants $\kA$ through $\kD$ are not well known, we have tried a number of different parameter settings in order to best match the model to our data.  The best effort result is given in Table \ref{tab:model} and the corresponding model curves are plotted as Curves 3 and 4 in Fig. \ref{fig:eprlaserpower}.
We find that the extended model visibly improves the agreement with the experiment, and that both charge state conversion processes must be included to obtain a satisfactory agreement.  The charge recombination rates that best match our experimental data are, however, much lower than the ones found in recent optical spectroscopy experiments \cite{jayakumar16}.
In addition, a significant discrepancy clearly remains for laser intensities above $10\unit{mW/mm^2}$.  We therefore conclude that the anomalous saturation cannot be fully accounted for by simple charge transfer dynamics between NV and P1 centers.
\begin{table}
\begin{tabular}[b]{lcc}
\hline\hline
Parameter                               & \quad\quad NV center \quad\quad  & \quad\quad P1 center \quad\quad \\
\hline 
Spin relaxation rates $\yNV$ and $\yN$  & $55\unit{s^{-1}}$     & $55\unit{s^{-1}}$  \\
Cross relaxation rate $\Delta$          & \multicolumn{2}{c}{$>10^5\unit{s^{-1}}$}  \\
Optical pump rate $k_0$                 & $0-160\unit{s^{-1}}$   & ---  \\
Ionization rate $k_A$ and $k_C$         & \multicolumn{2}{c}{$\ko$}  \\
Recombination rate $k_B$ and $k_D$      & \multicolumn{2}{c}{$10^3\unit{s^{-1}}$}  \\
Ratio of spin reservoirs                & \multicolumn{2}{c}{$R\approx 34$}  \\
\hline\hline
\end{tabular}
\caption{\captionstyle Parameters used for the model curves in Fig. \ref{fig:eprlaserpower}.  $\yNV$ and $\yN$ values were chosen such that the model best reproduced the low intensity ($I<4\unit{mW/m^2}$) part of the data.  We note that these values are in excellent agreement with reported rates at $200\unit{K}$ temperature \cite{reynhardt98,jarmola12}.  $\Delta$ is estimated $\sim 1\unit{MHz}$ based on the dipolar coupling between nearest-neighbor spins.  $k_0$ is according to the text.  $k_A$ and $k_C$ represent upper bounds \cite{manson05,bourgeois15}.
$k_B$ and $k_D$ represent the values that led to the best agreement between model and experiment.
$R$ is the ratio of effective densities in Table \ref{tab:enhancement}.  Parameters are for diamond chip A. 
}
\label{tab:model}
\end{table}

\section{Conclusions and Outlook}


While more work will be needed to explain the anomalous saturation behavior, we note that the effect could be accounted for by deep trap states formed by additional defects in the diamond crystals.  These trap states would act as sinks and slowly deprive NV and P1 centers of their electrons.  In order to be effective, the trap states would need to lie far below the conduction band so that they cannot be excited by green laser illumination.  There are several defects that could act as such trap states, such as di-vacancies formed during electron irradiation \cite{deak14}.
Additional evidence for a trap state hypothesis comes through recent photoconductivity measurement made by Chen \etal \cite{chen16}.  There, an unexpected quenching of pulsed photocurrents by low-intensity laser illumination was observed.  The authors suggested a similar trap state model that slowly depleted the $\No$ reservoir.  This would lead to a concurrent reduction of $\NV$.  Chen \etal\ also observed that short, intense laser pulses could replenish the $\No$ reservoir.  Pulsed laser excitation may therefore provide a means for also mitigating the anomalous saturation effect.

In conclusion, we have demonstrated a method for polarizing nitrogen donors in bulk diamond through optical pumping of NV centers and cross relaxation.  The polarization transfer between NV centers and nitrogen donors was found to be very efficient when operating at the energy level matching condition at 51 mT.  The maximum nitrogen donor polarization was 0.9 \% corresponding to a gain by 25 over the Boltzmann polarization.  The polarization enhancement saturated for an unexpectedly low illumination intensity of $\sim 10\unit{mW/mm^2}$, probably due to charge state conversion processes.  Overcoming this anomalous saturation will be crucial for realizing diamond polarizer devices for dynamic nuclear polarization.

The authors thank G. Jeschke for access to the X-band spectrometer.
T.F.S. acknowledges Society in Science, The Branco Weiss Fellowship, administered by the ETH Zurich.


\end{document}